\documentclass[sigconf]{acmart}

\AtBeginDocument{%
 }

\copyrightyear{2023}
\acmYear{2023}
\setcopyright{rightsretained}
\acmConference[MM '23]{Proceedings of the 31st ACM International Conference on Multimedia}{October 29-November 3, 2023}{Ottawa, ON, Canada}
\acmBooktitle{Proceedings of the 31st ACM International Conference on Multimedia (MM '23), October 29-November 3, 2023, Ottawa, ON, Canada}
\acmDOI{10.1145/3581783.3612526}
\acmISBN{979-8-4007-0108-5/23/10}

\makeatletter
\gdef\@copyrightpermission{
  \begin{minipage}{0.3\columnwidth}
   \href{https://creativecommons.org/licenses/by/4.0/}{\includegraphics[width=0.90\textwidth]{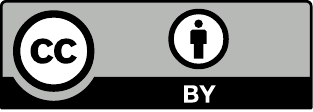}}
  \end{minipage}\hfill
  \begin{minipage}{0.7\columnwidth}
   \href{https://creativecommons.org/licenses/by/4.0/}{This work is licensed under a Creative Commons Attribution International 4.0 License.}
  \end{minipage}
  \vspace{5pt}
}
\makeatother

\usepackage{multirow}
\usepackage{makecell}
\usepackage{url}
\usepackage{subfigure}
\usepackage{soul}

\newcommand{\longname}{Cross-modality Augmented Multimedia Event Learning}
\newcommand{\shortname}{CAMEL}
\newcommand{\mtet}{M$^2$E$^2$}

\begin{document}

\title{Training Multimedia Event Extraction With Generated Images and Captions} 

\def\shorttitle{Training Multimedia Event Extraction With Synthetic Images and Captions}


\author{Zilin Du}
\email{zilin003@e.ntu.edu.sg}
\affiliation{
  \institution{Nanyang Technological University}
  \country{Singapore}
}

\author{Yunxin Li}
\email{liyunxin987@163.com}
\affiliation{
  \institution{Harbin Institute of Technology}
  \city{Shenzhen}
  \country{China}
}

\author{Xu Guo}
\email{xu008@e.ntu.edu.sg}
\affiliation{
  \institution{Nanyang Technological University}
  \country{Singapore}
}

\author{Yidan Sun}
\email{suny0053@e.ntu.edu.sg}
\affiliation{
  \institution{Nanyang Technological University}
  \country{Singapore}
}

\author{Boyang Li}
\email{boyang.li@ntu.edu.sg}
\affiliation{
    \institution{Nanyang Technological University}
  \country{Singapore}
  }


\begin{abstract}
Contemporary news reporting increasingly features multimedia content, motivating research on multimedia event extraction. 
However, the task lacks annotated multimodal training data and artificially generated training data suffer from distribution shift from real-world data. 
In this paper, we propose \longname{} (\shortname), which successfully utilizes artificially generated multimodal training data and achieves state-of-the-art performance. We start with two labeled unimodal datasets in text and image respectively, and generate the missing modality using off-the-shelf image generators like Stable Diffusion \cite{rombach2022high} and image captioners like BLIP \cite{li2022blip}. After that, we train the network on the resultant multimodal datasets. In order to learn robust features that are effective across domains, we devise an iterative and gradual training strategy. Substantial experiments show that \shortname{} surpasses state-of-the-art (SOTA) baselines on the \mtet{}  benchmark. On multimedia events in particular, we outperform the prior SOTA by 4.2\% F1 on event mention identification and by 9.8\% F1 on argument identification, which indicates that \shortname{} learns synergistic representations from the two modalities. Our work demonstrates a recipe to unleash the power of synthetic training data in structured prediction. 
\end{abstract}

\ccsdesc[500]{Information systems~Multimedia information systems}
\ccsdesc[500]{Computing methodologies~Structured outputs}

\keywords{Event Extraction; Multi-modal Learning; Data Augmentation; Cross-modality Generation}



\maketitle

\section{Introduction}


\begin{figure}[t]
  \centering
  \includegraphics[width=\linewidth]{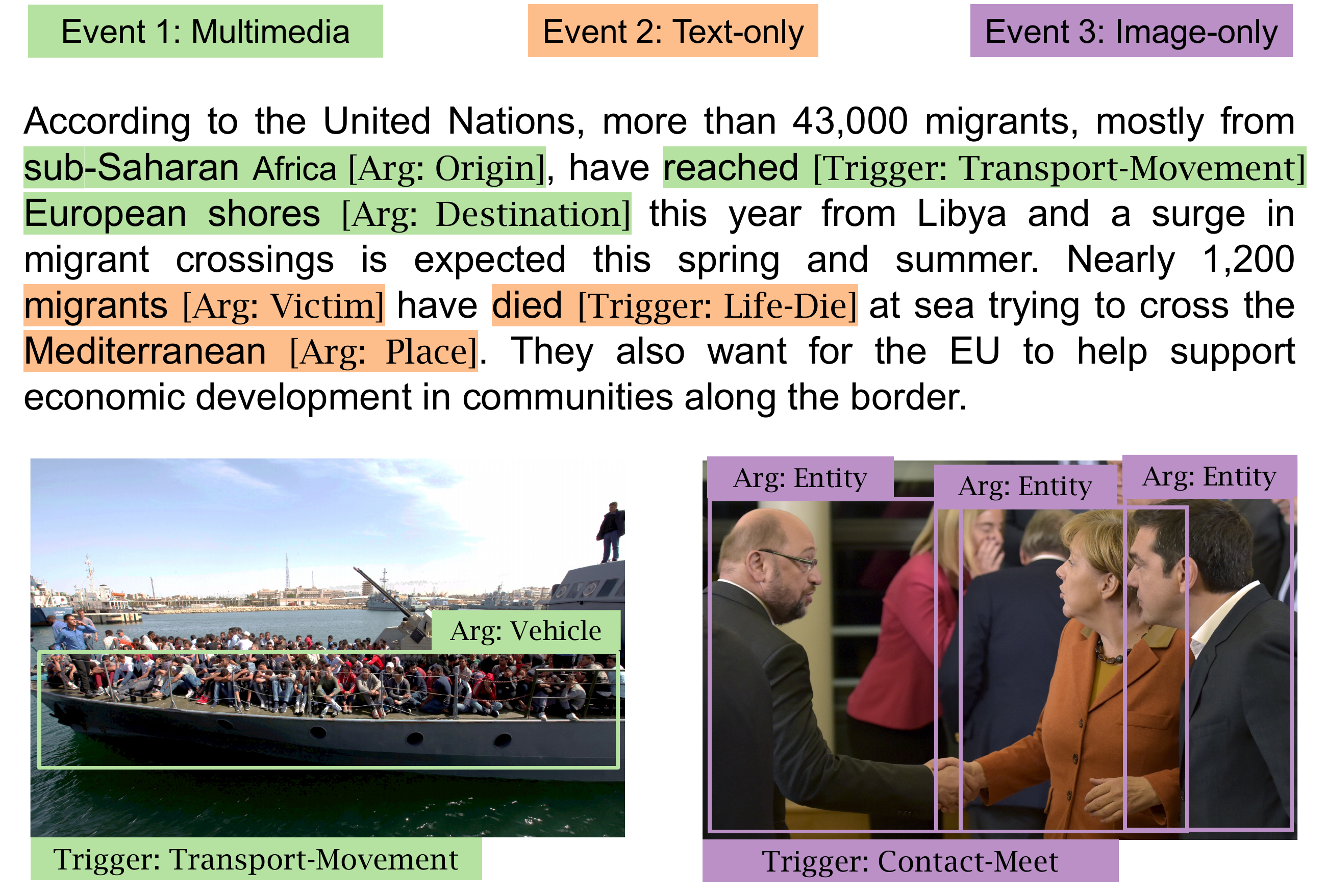}
  \caption{Multimedia Event Extraction: three events are extracted from a multimedia news article. The multimedia event (Green) Transport-Movement is triggered by the word `reached' and the image on the left. The textual event Life-Die (Orange) is triggered by the word `died' only, and the visual event Contact-Meet (Purple) is solely triggered by the image on the right.}
  \label{IMGexample}
\end{figure}

As a fundamental research topic in the domain of information extraction, event extraction aims to identify instances of events and their arguments from unstructured data \cite{doddington2004automatic,gildea2002automatic,kingsbury2002treebank,xue2003annotating,riloff1998empirical}. An event refers to a specific incident that involves a change in state, which are marked by triggers such as verbs. The arguments of an event include the time and place of the event occurrence and its participants, such as the initiator, the recipient, and the instrument. Traditional research mostly focuses on a single modality, either language  \cite{wang2021cleve, ma2022prompt, hsu2022degree, zhou2021role, du2021grit,li2022survey} or visual data \cite{yatskar2016situation, pratt2020grounded, cheng2022gsrformer, cho2022collaborative}. 

\begin{figure*}
  \centering
  \includegraphics[width=0.9\textwidth]{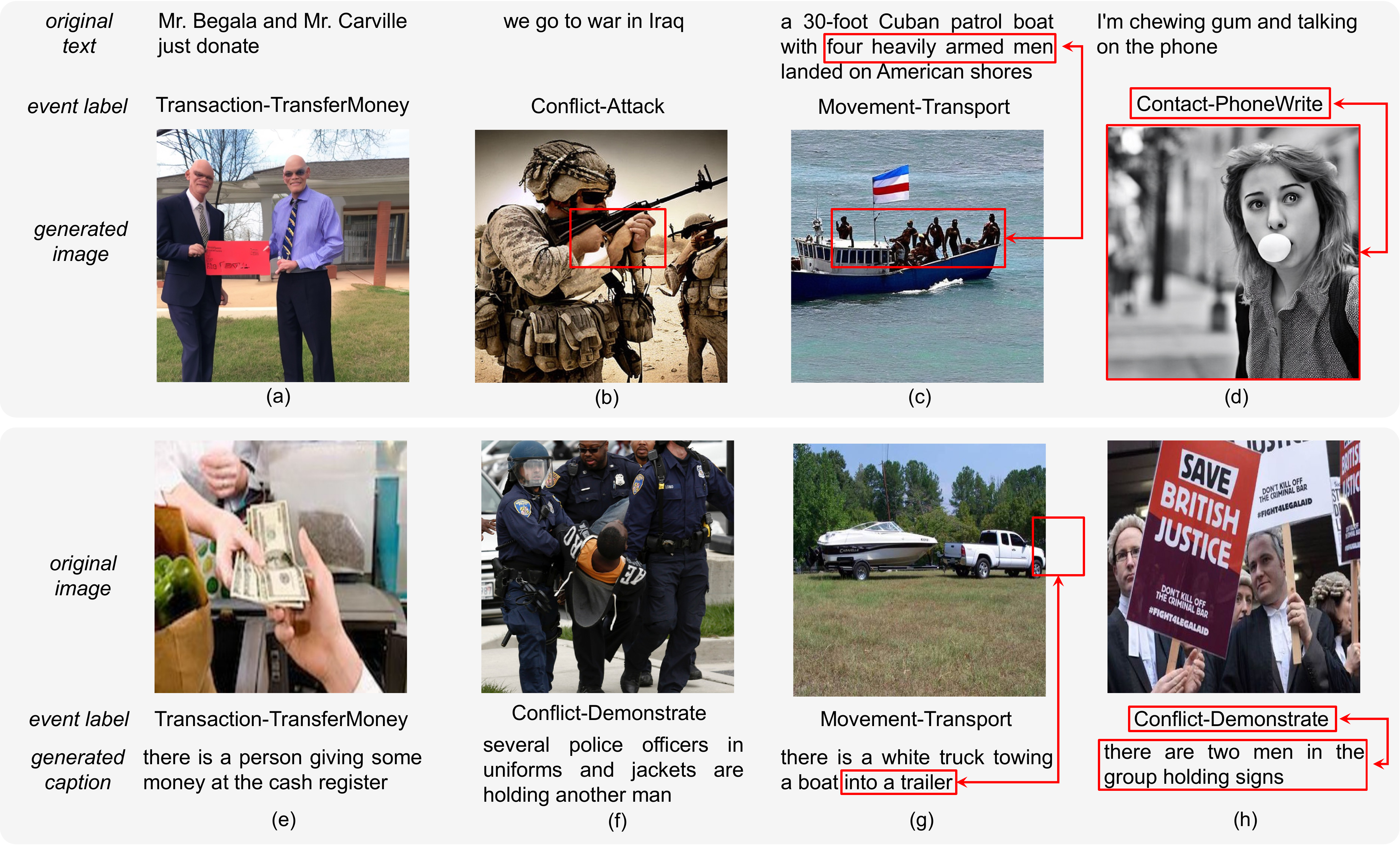}
  \caption{Examples of cross-modality augmented data. The red boxes indicate noise in the generated data, including inconsistency with the event label, hallucination, and unnatural image artifacts.} 
  \label{IMGexapmles}
\end{figure*}




As digital media quickly evolve, news reports today frequently present information with a combination of text and image, providing a more comprehensive view of events than text alone \cite{gao2020survey,wang2023reelframer}. 
This has spurred the emergence of the multimedia event extraction (MEE) task \cite{li2020cross}, which aims to jointly extract both textual and visual events from multimedia news articles. Figure \ref{IMGexample} shows an MEE instance. Interestingly, not all events in a multimedia news article are multimodal. For example, the event \texttt{Transport-Movement} is described by both the text and the image modalities, whereas the 
events \texttt{Life-Die} and \texttt{Contact-Meet} are contained respectively in text and image only. 

A major challenge posed by the MEE problem is the lack of multimodal training data. The \mtet{} dataset provided by \cite{li2020cross} is only the test set. The labeled training datasets ACE2005 \cite{walker2006ace} and imSitu \cite{yatskar2016situation} contain event annotations in a single modality only. Despite recent progress \cite{tong2020image,zhang2017improving}, transferring the knowledge learned from unimodal annotations to multimodal test data remains a difficult challenge. 

After the explosive success of image generation networks such as DALL-E 2 \cite{ramesh2021zero} and Stable Diffusion \cite{rombach2022high}, a natural thought is to perform cross-modality data augmentation in order to bridge the modality gaps of MEE. That is, conditioned on existing unimodal data, we can generate training data for the missing modality. After that, we use the resultant multimodal data to train a network. As the generative models capture world knowledge learned from observing correlative patterns among natural images and their textual descriptions, it is probable that such knowledge can be distilled and used to inform the task of event extraction.

However, a naive cross-modal data augmentation approach faces two obstacles. First, it is difficult to precisely control the generative models and produce data that are relevant to the event label and free of hallucination. In Figure \ref{IMGexapmles} (d), the generated image depicts gum chewing but the event label is about talking on the phone. In Figure \ref{IMGexapmles} (h), the generated caption describes the people as holding signs, whereas the label is demonstrating. In Figure \ref{IMGexapmles} (g), the caption hallucinates a trailer that does not exist in the image. Second, in the case of image generation, existing models occasionally still generate images with significant deformation and unnatural artifacts. For example, the soldier in Figure \ref{IMGexapmles} (b) is shown with three hands. For these reasons, the distribution of the generated data likely diverges from that of real-world data. In practice, we find that directly training on generated data results in performance degradation  (Table \ref{tab:abl_result}). 


To fully utilize the power of generative models to augment existing unimodal training data, we propose \longname{} (\shortname{}). After generating synthetic multimodal data, \shortname{} applies an iterative and gradual training strategy that learns robust representations under noisy data and distribution shifts. We train the networks using text coupled with synthetic visual data and images coupled with synthetic textual data. The network is gradually frozen from the bottom up during training. Experimentally, we show that this training technique offers substantial benefits over naive data augmentation. In particular, on multimedia events, we outperform the previous best network, UniCL \cite{liu2022multimedia}, by 4.1\% F1 on event mention identification and 9.8\% on argument role identification. 
In addition, the training strategy of \shortname{} works robustly under different choices of image generation and captioning networks.

Our contributions can be summarized as follows:
\begin{itemize}
\item For multimodal event extraction, \shortname{} utilizes synthetic data to fill in the missing modality in the unimodal ACE2005 and imSitu training datasets. To our best knowledge, this is the first work that successfully demonstrates the use of bi-directional cross-modality data augmentation (text-to-image and image-to-text) for multimodal learning. This results in superior data efficiency --- with the unlabeled real-world multimodal VOA dataset \cite{li2020cross} removed from training, we outperform previous work trained using VOA. 

\item We propose an incremental training strategy that handles artifacts, hallucination, and distribution shifts present in artificially generated multimodal data and avoids performance degradation caused by such noises. 

\item With \shortname, we set a new state of the art on the \mtet{} benchmark. On multimedia events in particular, we outperform the prior SOTA by 4.2\% F1 on event mention identification and by 9.8\% F1 on argument identification, which indicates that \shortname{} learns synergistic representations from multimodal data. 
\end{itemize}


\section{Related Work}
\label{Related Work}

\subsection{Event Extraction}

Event extraction \cite{li2022survey} is a well-studied problem in information extraction. Many early works \cite{liu2018jointly, wiedmann2017joint, wang2021cleve, ma2022prompt, hsu2022degree, zhou2021role, du2021grit} focus on textual data and aim to identify event structures containing trigger words and arguments from unstructured text. Traditionally, textual event extraction is formulated as sequence labeling \cite{ramponi2020biomedical, gui2020uncertainty, xu2021document, wen2021event}. More recent studies also formulate the problem as question answering \cite{zhou2021role, wei2021trigger, he2015question}. Similarly, visual event extraction \cite{yatskar2016situation, pratt2020grounded, wei2022rethinking, cho2022CoFormer,sadhu2021visual, sadhu2021video}, also referred to as situation recognition or visual semantic role labelling, aims to identify visual events and their participants. The earlier CRF-based methods \cite{yatskar2016situation, yatskar2017commonly} jointly predict event type and the associated roles in one stage. \cite{mallya2017recurrent} shows that identifying the action and the argument roles in two stages with separate networks to offer performance gains. More recent methods, such as GSRFormer \cite{cheng2022gsrformer} and SituFormer \cite{wei2022rethinking}, adopt the two-stage approach.

Several studies investigate the use of multimodal data in unimodal event extraction. For example, \cite{tong2020image} and \cite{zhang2017improving} retrieve image relevant to the events, which can assist with disambiguation.  \cite{li2022clip} leverages image captions as distant supervision to interpret events in the associated images. Although they operate on multimodal data, these methods are aimed at events present in one modality. 

Multimedia event extraction is proposed to extract events and arguments from multimedia documents \cite{li2020cross, chen2021joint}. \cite{li2020cross} tackles image-text documents while \cite{chen2021joint} focuses on video. WASE \cite{li2020cross} uses weakly supervised learning to encode structured representations from textual and visual data into a shared embedding space. \cite{liu2022multimedia} introduces contrastive learning to bridge textual and visual modalities. Compare to these research, our method is the first to directly learn from synthetic multimodal training data, which are generated from labeled unimodal data.

\subsection{Cross-modality Generation}

\subsubsection{Cross-modality Generative Models} 
Text-to-image and image-to-text generative models are gaining traction. Text-to-image models \cite{reed2016generative, li2019controllable, yu2021vector, ramesh2021zero, yu2022scaling, croitoru2023diffusion} are developed to produce high-quality images based on natural language descriptions. Early studies are based on GANs \cite{reed2016generative, li2019controllable, yu2021vector}, auto-regression \cite{ramesh2021zero, yu2022scaling}, and VAEs \cite{higgins2016beta, vahdat2020nvae}. More recently, diffusion models have achieved impressive results \cite{zhang2023text, croitoru2023diffusion}. Some, like GLIDE \cite{nichol2022glide} and Imagen \cite{saharia2022photorealistic}, generate images at the pixel level directly; others, like DALL-E 2 \cite{saharia2022photorealistic} and Stable
Diffusion \cite{rombach2022high}, operate on a low-dimensional latent space. They have shown great promise in high-fidelity image generation. Meanwhile, remarkable improvement has also been achieved in the field of image-to-text generation (a.k.a. image captioning) with models like BLIP \cite{li2022blip}, GiT \cite{wang2022git}, and so on \cite{li2023blip, li2023lmeye, kumar2022imagecaptioning, wang2022ofa}. 

\subsubsection{Cross-modality Generative Data Augmentation}

Recent advances in generative models have propelled data augmentation research to a new level. On textual tasks, one approach is to generate additional textual training data \cite{west2022symbolic, wang2022self, honovich2022unnatural, meng2022generating, gao2023self, ye2022zerogen, meng2022tuning}. Another, multimodal approach is to generate visual data to complement existing textual data \cite{yang2022z, lu2022imagination, zhu2022visualize, long2021generative}, which improves performance on textual tasks. For example, \cite{long2021generative} generates visual data for machine translation. \cite{zhu2022visualize} uses generated images to guide text generation tasks, such as text completion, story generation, and concept-to-text generation. In addition, \cite{yang2022z,lu2022imagination} integrate synthetic images into language models to enhance the solution of plain language understanding tasks under low-resource settings. Unlike previous studies that address unimodal problems by synthesizing multimodal data, our work use the generated data to tackle multi-modal tasks. Doing so places a stringent requirement on the quality of generated data, as we need to train encoders in both modalities with the generated data. This necessitates overcoming the domain shifts between generated and real data. To the best of our knowledge, this is the first work to utilize bidirectional cross-modality data generation models for multimodal tasks.

\section{Task Definition}
\label{Definition}

Let $D=\langle M, S\rangle$ represent a multimedia document, which consists of a set of images $M = \{m_1,m_2,\ldots \}$ and a set of sentences $S = \{s_1,s_2,\ldots \}$. Each sentence $s$ consists of a sequence of words $[w_1, w_2, \ldots, w_L]$. The multimedia event extraction task contains the following two components.

\textbf{Event Mention Identification:} Given a multimedia document $D$, the first goal is to identify a set of event mentions from $D$. An event mention $e$ belongs to one of the predefined event types, $y_e$, and is grounded on a trigger word $w$ or a trigger image $m$ or both. A multimedia event contains both a trigger word $w$ and trigger image $m$, while a text-only or an image-only event only contains one type of trigger.

\textbf{Argument Role Identification:} The purpose of argument role identification is to extract, from the document $D$, all participants and attributes (i.e., arguments) of a given event $e$. For each event type, there is a predefined list of argument types. Each argument $a$ is classified into one argument type $y_a$ associated with the event type. The argument is grounded on a textual span $t$ in a sentence or one or more object bounding boxes in the image. The algorithm for argument role identification must also identify the position of the textual span $t$ and the bounding boxes. 

If $e$ is a multimedia event, it must be grounded on both a textual trigger and a visual trigger. The arguments of multimedia events could contain both textual spans and visual objects. For example, in Figure \ref{IMGexample}, the multimedia event \texttt{Transport: Movement} is grounded on both the trigger word ``reached'' and the trigger image on the left. It also has two textual arguments and one visual argument.

\section{Approach}
\label{Methdology}

The proposed approach, \shortname{}, is trained with multimedia data that are artificially generated from unimodal data (Section \ref{DataDis}). The cross-modality generative data augmentation approach can be thought of as distilling event-related knowledge from large generative models to the event identification network. 

\begin{figure*}
  \centering
  \includegraphics[width=0.9\textwidth]{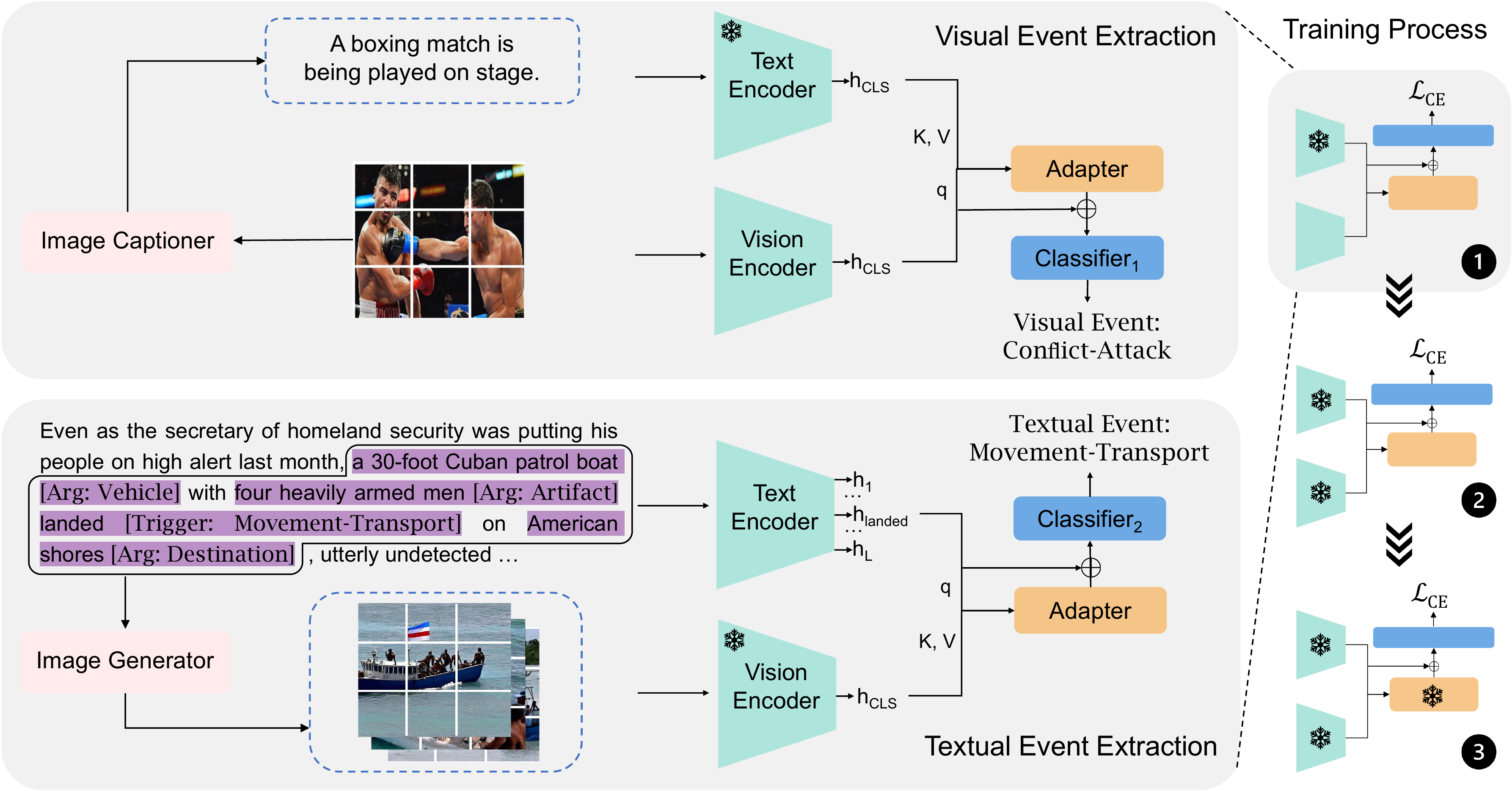}
  \caption{An overview of the \shortname{} network architecture and its training strategy}
  \label{IMGarchitecture}
\end{figure*}

We show an overview of \shortname{} in Figure \ref{IMGarchitecture}. In a dual-encoder architecture, \shortname{} first extracts features from the two modalities separately using unimodal encoders. To perform feature fusion and allow the network to pick relevant features among possibly noisy input, we design a modality-shared adapter module that perform cross-attention between the modalities. Further, to cope with possible distribution shifts and learn robust and generalizable features, we employ an iterative and gradual training strategy (Section \ref{training}). After these steps, we feed the resultant representation to domain-specific classifiers to identify the event mentions and arguments.

\subsection{Cross-modality Generative Data Augmentation}
\label{DataDis}

A major obstacle for multimedia event extraction is the lack of multimodal training data. In the commonly used setup, first proposed by \cite{li2020cross}, the training data contains event annotations on text (ACE2005 \cite{walker2006ace}) and event annotations on images (imSitu \cite{yatskar2016situation}). The unlabeled VOA \cite{li2020cross} dataset is often used as auxiliary training data;it contains parallel image-text data but no event annotations. 

To tackle this problem, we utilize large text-to-image and image-to-text generative models to perform cross-modality generative data augmentation. Specifically, to augment the labeled textual data, we generate images using a text-to-image model. In addition, to augment labeled image data, we use an image-to-text model to generate image captions. This procedure yields labeled parallel image-text data. For most of our experiments, we use Stable Diffusion v2.1 \cite{rombach2022high} for image generation and BLIP \cite{li2022blip} for captioning. However, \shortname{} can be applied to a range of generative models with little loss in performance, as demonstrated in Section \ref{sec:ablation}.

\vspace{0.1in}
\noindent 
\textbf{Visual Data Augmentation.} 
We perform visual data augmentation on the labelled textual dataset, ACE2005, which consists of textual news reports. In order to extract textual spans that are relevant to the event, we utilize the existing annotations of event arguments and trigger words. For each event, we find the shortest continuous textual span that include all arguments and the trigger word, and use that as the textual input to image generation networks. We show one example of the extracted text span in the purple-lined box in Figure \ref{IMGarchitecture}. 

The image generation process is stochastic. Thus, we generate several images for each textual event in ACE2005 to cover different possible visual appearances and spatial arrangements. The number of images is a hyperparameter, which we set to four. 


\vspace{0.1in}
\noindent 
\textbf{Textual Data Augmentation.} To augment the visual dataset imSitu with the textual modality, we utilize the off-the-shelf image-to-text model to generate image captions. To generate diverse and detailed captions, we adopt nucleus sampling \cite{holtzman2019curious}. At each time step, the technique iteratively adds the most probable word to the candidate list until the total probability of the candidates exceeds a pre-defined probability. After that, the probabilities of candidates are normalized and one word is sampled accordingly. 
We generate one caption for each image in imSitu.

\subsection{Model Architecture}

\vspace{0.1in}
\noindent 
\textbf{Feature Extractors.}
\shortname{} utilizes two pretrained Transformer encoders to extract unimodal features separately. Using the hidden states of the last network layer, the text encoder obtains a $d$-dimensional vector representation $h^{\text{text}}_i$ for each word $w_i$. Similarly, each patch of the image is encoded into a $d$-dimensional vector $h^{\text{img}}_i$.
We denote the set of all text representations as $H_T$ and the set of all visual representations as $H_V$.  
We also prepend CLS tokens to the input of the two encoders. The corresponding encodings $h^{\text{text}}_\text{CLS}$ and $h^{\text{img}}_\text{CLS}$ can be understood as representing information from the entire sentence or image. 



\begin{figure}[t]
  \centering
  \includegraphics[width=0.8\linewidth]{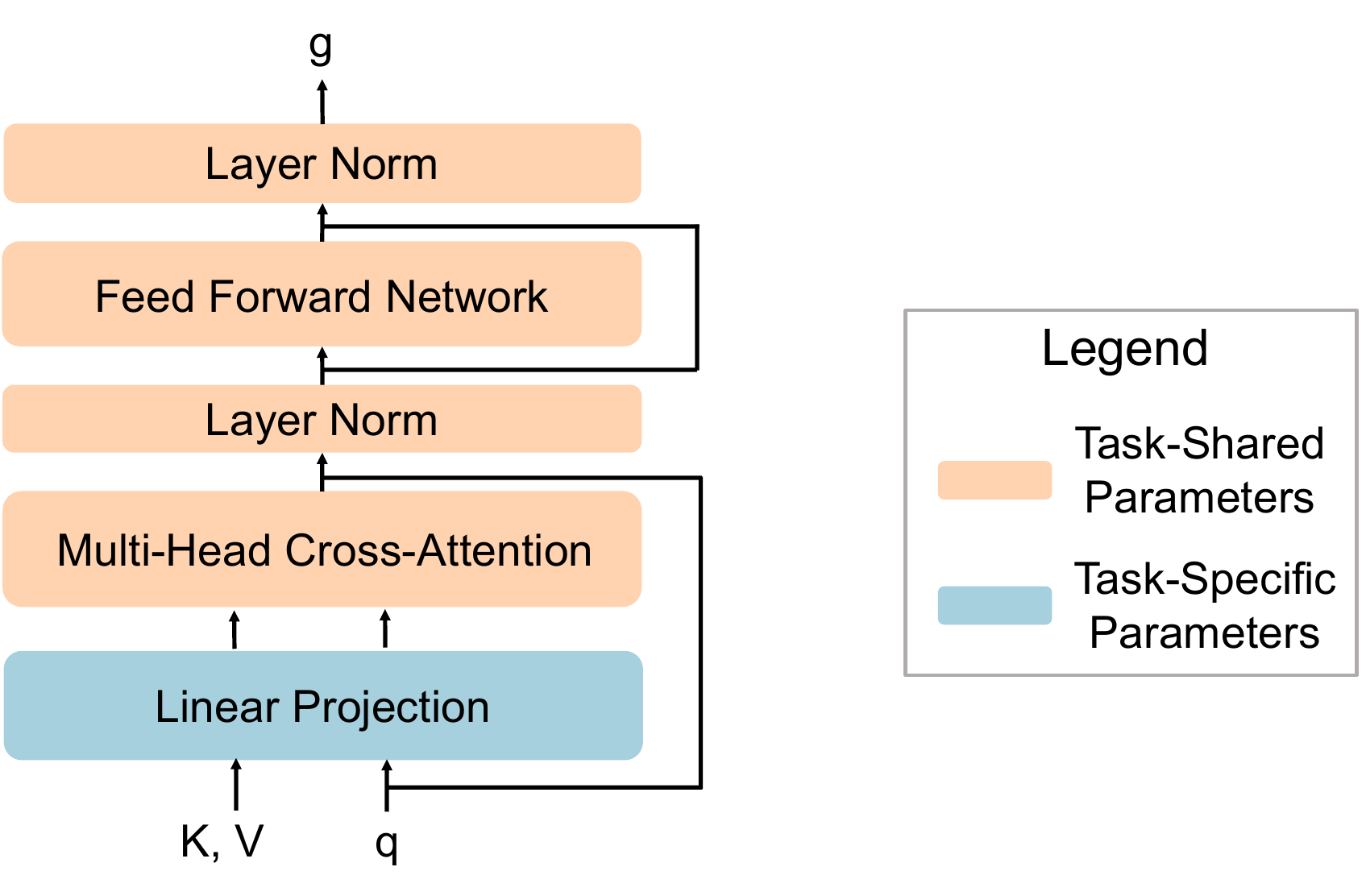}
  \caption{The architecture of the Adapter network.}
  \label{IMGAdapter}
\end{figure}

\vspace{0.1in}
\noindent 
\textbf{Feature Fusion.}
We devise a cross-attention network module, commonly used in Transformer networks, to fuse textual and visual features. The network consists of multi-head cross attention, layer normalization, and some linear layers. The detailed architecture is shown in Figure \ref{IMGAdapter}. 

We refer to this module as the Adapter network. For simplicity, we denote the input to the Adapter as the query vector $q$, the key matrix $K$, and the value matrix $V$. The overall network is denoted as the function 
\begin{equation}
g = \text{Adapter}(q, K, V). 
\end{equation}
We make repeated use of the same Adapter module with in the identification of event mentions and arguments, but change the $q$, $K$, and $V$ depending on the exact task. Most parameters are shared across tasks. However, parameters in the linear task-specific projection layer are specific to the four tasks (textual event mention, textual argument role, visual event mention, visual argument role).

The design of the Adapter module is motivated by the characteristics of multimedia documents, which usually do not explicit indicate the correspondence between images and the main text. When we try to identify a textual event and its arguments, we do not know which image is relevant to this event. The cross-attention mechanism allows the network to distinguish relevant images. Similarly, when extracting visual events, the network relies on the Adapter to select relevant portions of the text to facilitate its prediction.

\vspace{0.1in}
\noindent 
\textbf{Textual Event Extraction.}
The first sub-problem in textual event extraction is to identify the trigger word. This is word-level classification. The trigger word should be classified into the exact event type, whereas other words should be classified as non-triggers. 

For the classification of the $i^{\text{th}}$ word, we first take its encoding from the textual encoder, $h^{\text{text}}_i$. After that, we feed $h^{\text{text}}_i$ to the Adapter network as the query vector. We use the CLS token encodings of all images in the entire multimedia document, denoted as $H^{\text{all-img}}$, as $K$ and $V$ in cross attention.
\begin{equation}
g^{\text{text}}_i = \text{Adapter}(h^{\text{text}}_i, H^{\text{all-img}}, H^{\text{all-img}}).
\end{equation}
After that, we concatenate $h^{\text{text}}_i$ and $g^{\text{text}}_i$ and feed them through a linear classifier. The loss is cross-entropy. 

The second sub-problem is the identification of event arguments. Following the convention in the literature \cite{li2020cross,liu2022multimedia}, we use the ground-truth list of entities for both training and inference. Each entity is a textual span that describes a person, an organization, a location and so on. We take the encoding of the first word in that entity as the entity feature $h^{\text{text-ent}}$, and feed it to the Adapter. \begin{equation}
g^{\text{text-ent}}_i = \text{Adapter}(h^{\text{text-ent}}, H^{\text{all-img}}, H^{\text{all-img}}).
\end{equation}
Similarly, we concatenate $h^{\text{text-ent}}$, $g^{\text{text-ent}}_i$, and the textual encoding of the trigger word, and feed them through a linear classifier, which classifies it into the argument classes. Though the types of valid arguments change depending on the event, here we do not exploit this fact for further performance improvement. 

During training, if the ACE2005 sentence contains an event, we generate several positive images from the event text prompt (see Section \ref{DataDis}). In addition, we also include some negative images generated for other events into $H^{\text{all-img}}$. This trains the network to distinguish between relevant images and irrelevant images. However, if the ACE2005 sentence does not contain any event, we would not be able to extract the event prompt using the method in Section \ref{DataDis}. Instead, we randomly sample generated images from other text and use their encodings as $H^{\text{all-img}}$.

\vspace{0.1in}
\noindent 
\textbf{Visual Event Extraction.}
Similar to the textual modality, visual event extraction has two sub-problems, the classification of images into event types or non-events, and identification of objects as event arguments. For image event classification, we take the encoding of the image CLS token, $h^{\text{img}}_\text{CLS}$. Using the Adapter network again, we acquire an aggregated feature from the text modality, which we denote as $g^{\text{img}}$,
\begin{equation}
g^{\text{img}} = \text{Adapter}(h^{\text{img}}_\text{CLS}, H^{\text{all-text}}, H^{\text{all-text}}),
\end{equation}
where the matrix $H^{\text{all-text}}$ contains the encoding vectors of the textual CLS token encodings of all sentences in the same batch. We feed the concatenation of $h^{\text{img}}_\text{CLS}$ and $g^{\text{img}}$ to a linear classifier. 

\begin{figure}[t]
  \centering
  \includegraphics[width=\linewidth]{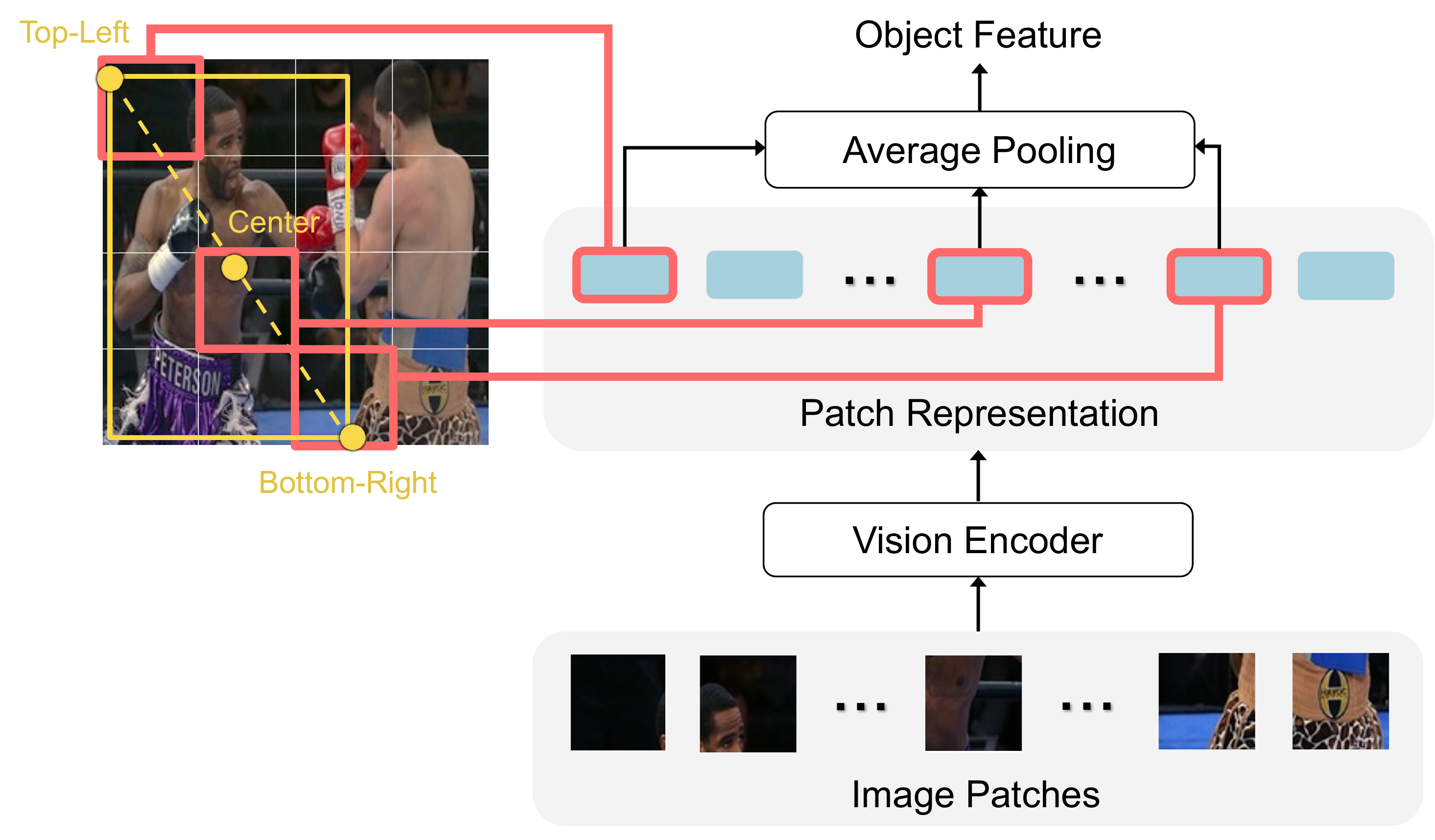}
  \caption{Extracting features for objects in images.}
  \label{IMGObject}
\end{figure}

For the second sub-problem, event argument identification, we first extract all objects in an image using an off-the-shelf object detector. For each object bounding box, we identify the three patches that contain its top-left corner, its center, and its bottom-right corner respectively. After that, we take the average of the three patch encodings, which we denote as $h^{\text{img-obj}}$. The object feature extraction process is illustrated in Figure \ref{IMGObject}. Once again, we apply feature fusion using the adapter network to obtain $g^{\text{img-obj}}$, 
\begin{equation}
g^{\text{img-obj}} = \text{Adapter}(h^{\text{img-obj}}, H^{\text{all-text}}, H^{\text{all-text}}).
\end{equation}
Finally, we concatenate three feature vectors, $h^{\text{img-obj}}$, $g^{\text{img-obj}}$, and $h^{\text{img}}$ into a single vector and feed it through a linear classifier.

\subsection{Multimedia Event Extraction}
For multimedia events, we need to resolve the coreference between text events and visual events. Given a multimedia document, we compute the similarity of each sentence-image pair. Following \cite{li2020cross, liu2022multimedia}, we treat a textual event and a visual event as the same event if and only if they have the same event type and the similarity between the sentence and image is greater than a threshold. We calculate the cosine similarity of the sentence-image pair using the CLIP model \cite{radford2021learning}. The multimedia event inherits all textual event arguments and visual event arguments as its own arguments.

\subsection{Training Strategy}
\label{training}

\sloppypar{
Robust representation learning is key to the success of cross-modality data augmentation and multimedia event extraction. 
As discussed in the introduction and shown in Figure \ref{IMGexample}, the automatically generated multimodal data often contain noise, such as inconsistency with the event label, hallucination, unnatural image artifacts, and so on. The discrepancy between the generated data distribution and the real-world data distribution may cause generalization difficulties. 
In addition, the \mtet{} task itself poses a transfer learning problem because the training data, ACE2005 and imSitu, have different distributions from the test set. Hence, we need to learn robust feature representations that generalize well.}

We propose an iterative and gradual training strategy, shown in the right column of Figure \ref{IMGarchitecture}. We divide the training into the three stages. In the first stage, we first train on visual event mention, followed by textual event mention. The separation is a simple method to alleviate the well-known problem that different modalities learn at different speeds \cite{Wang_2020_CVPR:multimodal-hard}. In the first stage, all network parameters are trained except the feature extractor corresponding to the generated synthetic data. For example, when training on real text data and generated image data, the text encoder is trainable but the image encoder is frozen. The rationale is to prevent the gigantic feature extractors (with hundreds of millions of parameters) from overfitting the low-level feature distributions of the augmented training data, which are likely idiosyncratic (e.g., soldiers with three hands) and not generalizable. However, we postulate that the high-level features extracted by the encoders are not heavily affected by shifts in lower-level feature distributions, so we train all the parameters after the encoders. 

In the second stage, we again train the network on visual event mention identification, followed by textual event mention identification. Both encoders are frozen and only the Adapter and classifiers are trained. The design rationale is to allow visual classifiers to adapt to changes in the textual encoder in the first stage, and vice versa. In the third stage, we freeze all network parameters but finetune the visual event mention classifier using balanced event data. This technique is to mitigate the negative effects of imbalanced class proportions in the visual event mentions \cite{kang2019decoupling}. Finally, we separately finetune the visual encoder for visual event argument identification, and fintune the text encoder for textual event argument identification. This creates two models specialized for argument identification.



\section{Experiments}
\label{Experiments}

In this section, we extensively evaluate  \shortname{} by comparing against existing SOTA approaches, against ablated version of \shortname{}, and against different choices for the image generators and image captioning networks. 

\subsection{Experimental Setting}

\vspace{0.1in}
\noindent
\textbf{Datasets and Evaluation.}
We evaluate on the \mtet{} benchmark, a large-scale multimedia event extraction dataset that with the 8 types of events and 15 types of arguments. It contains 245 multimedia documents with 6,167 sentences and 1,014 images. There are 1,297 textual events and 391 visual events, among of which 192 textual event mentions and 203 visual event mentions are aligned into 309 multimedia events. 

Since \mtet{} does not provide training data, we follow the previous work \cite{li2020cross, liu2022multimedia} to use the ACE2005 \cite{walker2006ace} and imSitu \cite{yatskar2016situation} (with the grounding information from \cite{pratt2020grounded}) for training. ACE2005 is a text dataset annotated with 33 event types, which contains the 8 specific types in \mtet{}. The image dataset imSitu is annotated with 504 activity verbs and 1,788 semantic roles. To utilize this dataset for 8-class classification, we follow \cite{li2020cross} and map the 98 activity verbs to the 8 event types of \mtet. Following the previous works on event extraction \cite{li2020cross, liu2018jointly, wiedmann2017joint}, we use precision (P), recall (R), and F1 score (F1) as the default evaluation metrics.

\vspace{0.1in}
\noindent
\textbf{Baselines.}
Following \cite{liu2022multimedia}, we compare \shortname{} with eight baselines for multimodal or unimodal event extraction. 

Multimodal event extraction techniques can extract both textual events and visual events. WASE \cite{li2020cross} first trains on different modalities independently and uses weakly supervised learning to align the two modalities. Two variations exist: WASE$_{\rm att}$ locates the visual arguments using an attention heat map, whereas WASE$_{\rm obj}$ leverages a object detection model. Flat$_{\rm att}$ and Flat$_{\rm obj}$ \cite{li2020cross} are the simplified versions of WASE$_{\rm att}$ and WASE$_{\rm obj}$ respectively; they remove the graph convolution networks and concatenate features of different modalities for classification. UniCL \cite{liu2022multimedia} is the state-of-the-art on \mtet{}, which incorporates visual knowledge into textual event extraction but uses two separate modality-specific models for event extraction.

Unimodal event extraction methods only extract textual or visual events but not both. JMEE \cite{liu2018jointly} is a state-of-the-art textual event extraction technique which utilizes an attention-based Graph Convolution Network. GAIL \cite{zhang2018event} is a reinforcement learning method for textual event extraction where rewards are estimated by a Generative Adversarial Network. VAD \cite{zhang2017improving} augments textual documents with images retrieved from the Internet to improve textual event extraction. Clip-Event \cite{li2022clip} utilizes the pretrained CLIP network to perform visual event extraction. WASE-T and WASE-V are the WASE model which trained on ACE2005 and imSitu only. The latter has two further varations WASE-V$_{\rm att}$ and WASE-V$_{\rm obj}$ \cite{li2020cross}.

\vspace{0.1in}
\noindent
\textbf{Hyperparameters. }
During cross-modality data augmentation, for each event in ACE20005, we perform one-time generation of 4 images at 512×512 resolution with 100 denoising steps. In addition, we use nucleus (top-$p$) sampling \cite{holtzman2019curious} for image captioning with a probability threshold $p$ of 0.9. We generate one caption for each original image.

For fair comparisons with the SOTA baseline \cite{liu2022multimedia}, we use the same 12-layer BERT$_{\text{Large}}$ as the text encoder, and the same 12-layer Transformer CLIP model \cite{jia2021scaling} as the visual encoder with 16x16 patch size. To detect objects for visual argument roles, we leverage the pretrained YOLOv8 \cite{Yolov8} as the object detector. 

When training on visual event extraction, our batch size is set to 64 and learning rate to $10^{-4}$. For textual event extraction, the batch size is set to 10 and  learning rate set to $10^{-4}$. We employ the AdamW optimizer \cite{Loshchilov2017DecoupledWD} with $10^{-2}$ weight decay coefficient and the cosine learning rate schedule. 
In the first round of training, we train on the visual modality for 10 epochs and on the textual modality for 5 epochs. In the remainder of training, only one epoch is used for any modality. The maximum text input length is 200. 

\subsection{Main Results}

\begin{table}
\centering
\small
  \caption{Main results on event mention and argument role extraction for three types of events.}
  \label{tab:main_result}
  \begin{tabular}{c c ccc|ccc}
    \toprule
     & & \multicolumn{3}{c}{\textbf{Event Mention}} & \multicolumn{3}{c}{\textbf{Argument Role}}\\
    Event & Method & P & R & F1 & P & R & F1\\
    \midrule
    \multirow{10}*{Textual} & JMEE \cite{liu2018jointly} & 42.5 & 58.2 & 48.7 & 22.9 & 28.3 & 25.3 \\
    ~ & GAIL \cite{zhang2018event} & 43.4 & 53.5 & 47.9 & 23.6 & 29.2 & 26.1 \\ 
    ~ & VAD \cite{zhang2017improving} & 34.8 & 64.4 & 45.2 & 23.1 & 27.5 & 25.1\\
    ~ & Flat \cite{li2020cross} & 34.2 & 63.2 & 44.4 & 20.1 & 27.1 & 23.1 \\  
    ~ & WASE-T \cite{li2020cross} & 42.3 & 58.4 & 48.2 & 21.4 & 30.1 & 24.9 \\
    ~ & WASE$_{\rm att}$ \cite{li2020cross} & 37.6 & 66.8 & 48.1 & 27.5 & 33.2 & 30.1\\ 
    ~ & WASE$_{\rm obj}$ \cite{li2020cross} & 42.8 & 61.9 & 50.6 & 23.5 & 30.3 & 26.4\\ 
    ~ & UniCL \cite{liu2022multimedia} & 49.1 & 59.2 & 53.7 & 27.8 & 34.3 & 30.7\\
    ~ & \shortname{} (Ours) & 45.1 & 71.8 & \textbf{55.4} & 24.8 & 41.8 & \textbf{31.1}\\ 
    \hline
    \multirow{9}*{Visual} & Flat\cite{li2020cross} & 27.1 & 57.3 & 36.7 & 4.3 & 8.9 & 5.8 \\
    ~ & WASE-V$_{\rm att}$ \cite{li2020cross} & 29.7 & 61.9 & 40.1 & 9.1 & 10.2 & 9.6 \\
    ~ & WASE-V$_{\rm obj}$ \cite{li2020cross} & 28.6 & 59.2 & 38.7 & 13.3 & 9.8 & 11.2 \\
    ~ & WASE$_{\rm att}$ \cite{li2020cross} & 32.3 & 63.4 & 42.8 & 9.7 & 11.1 & 10.3\\  
    ~ & WASE$_{\rm obj}$ \cite{li2020cross} & 43.1 & 59.2 & 49.9 & 14.5 & 10.1 & 11.9\\
    ~ & CLIP-Event \cite{li2022clip} & 41.3 & 72.8 & 52.7 & 21.1 & 13.1 & 17.1\\
    ~ & UniCL \cite{liu2022multimedia} & 54.6 & 60.9 & 57.6 & 16.9 & 13.8 & 15.2\\
    ~ & \shortname{} (Ours) & 52.1 & 66.8 & \textbf{58.5} & 21.4 & 28.4 & \textbf{24.4}\\ 
    \hline
    \multirow{5}*{Multi.} & Flat \cite{li2020cross} & 33.9 & 59.8 & 42.2 & 12.9 & 17.6 & 14.9 \\ 
    ~ & WASE$_{\rm att}$ \cite{li2020cross} & 38.2 & 67.1 & 49.1 & 18.6 & 21.6 & 19.9\\ 
    ~ & WASE$_{\rm obj}$ \cite{li2020cross} & 43.0 & 62.1 & 50.8 & 19.5 & 18.9 & 19.2 \\   
    ~ & UniCL \cite{liu2022multimedia} & 44.1 & 67.7 & 53.4 & 24.3 & 22.6 & 23.4\\
    ~ & \shortname{} (Ours) & 55.6 & 59.5 & \textbf{57.5} & 31.4 & 35.1 & \textbf{33.2} \\  
  \bottomrule
\end{tabular}
\end{table}

Table \ref{tab:main_result} presents the performance of our proposed method \shortname{} and several state-of-the-art baselines. The results show \shortname{} significantly improves the event extraction performance over baseline methods. On textual events, we surpass UniCL by 1.7\% F1 for event mention and 0.4\% F1 for argument role. On visual events, we surpass UniCL by 0.9\% F1 for event mention and 9.2\% F1 for argument role. We speculate that the relatively small improvements for textual argument roles is that some textual arguments are pronouns (e.g., she) or proper noun (e.g., Saudi Arabia), which are not straightforward to visualize by the image generators. 

Interestingly, the biggest performance boost appears on multimedia event extraction. We outperform the prior SOTA by 4.2\% F1 on event mention identification and by 9.8\% F1 on argument identification. This suggests \shortname{} effectively  learns synergistic representations from the two modalities. 


\subsection{Ablation Study}
\label{sec:ablation}

\begin{table*}
\centering
  \caption{Ablation results of \shortname{} on the \mtet{} dataset. }
  \small
  \label{tab:abl_result}
  \begin{tabular}{c ccc|ccc|ccc|ccc|ccc|ccc}
    \toprule
     & \multicolumn{6}{c}{\textbf{Textual Events}} & \multicolumn{6}{c}{\textbf{Visual Events}} & \multicolumn{6}{c}{\textbf{Multimedia Events}}\\
     & \multicolumn{3}{c}{\textbf{Event Mention}} & \multicolumn{3}{c}{\textbf{Argument Role}}& \multicolumn{3}{c}{\textbf{Event Mention}} & \multicolumn{3}{c}{\textbf{Argument Role}}& \multicolumn{3}{c}{\textbf{Event Mention}} & \multicolumn{3}{c}{\textbf{Argument Role}}\\
    Method & P & R & F1 & P & R & F1& P & R & F1 & P & R & F1& P & R & F1 & P & R & F1\\
    \midrule
    
    \shortname{} & 45.1 & 71.8 & \textbf{55.4} & 24.8 & 41.8 & \textbf{31.1} & 52.1 & 66.8 & \textbf{58.5} & 21.4 & 28.4 & \textbf{24.4} & 55.6 & 59.5 & \textbf{57.5} & 31.4 & 35.1 & \textbf{33.2} \\
    combined training & 41.9 & 69.7 & 48.3 & 22.0 & 34.5 & 26.8 & 60.1 & 40.4 & 48.3 & 24.8 & 17.7 & 20.6 & 53.2 & 32.0 & 40.0 & 27.4 & 15.8 & 20.0 \\ 
    one-round training & 45.1 & 70.6 & 55.0 & 22.4 & 40.6 & 30.6 & 66.5 & 36.6 & 47.2 & 24.1 & 13.5 & 17.3 & 55.9 & 33.6 & 42.0 & 31.5 & 19.2 & 23.8\\
    w/o augmentation & 40.0 & 73.2 & 51.7 & 25.7 & 30.5 & 27.9 & 48.9 & 62.9 & 55.0 & 19.3 & 25.9 & 22.1 & 51.5 & 54.4 & 52.9 & 31.6 & 26.4 & 28.8\\
    w/o adapter & 43.7 & 70.8 & 54.0 & 25.3 & 36.0 & 29.7 & 45.5 & 68.0 & 54.5 & 19.3 & 30.5 & 23.6 & 49.8 & 57.0 & 53.2 & 30.0 & 30.9 & 30.4\\
    
  \bottomrule
\end{tabular}
\end{table*}

\begin{table*}[!t]
\centering
  \caption{Performance of \shortname{} with different image captioners and image generators.}
  \small
  \label{tab:gen}
  \begin{tabular}{@{}c ccc|ccc|ccc|ccc|ccc|ccc@{}}
    \toprule
     & \multicolumn{6}{c}{\textbf{Textual Events}} & \multicolumn{6}{c}{\textbf{Visual Events}} & \multicolumn{6}{c}{\textbf{Multimedia Events}}\\
     & \multicolumn{3}{c}{\textbf{Event Mention}} & \multicolumn{3}{c}{\textbf{Argument Role}}& \multicolumn{3}{c}{\textbf{Event Mention}} & \multicolumn{3}{c}{\textbf{Argument Role}}& \multicolumn{3}{c}{\textbf{Event Mention}} & \multicolumn{3}{c}{\textbf{Argument Role}}\\
    \midrule
    Method & P & R & F1 & P & R & F1& P & R & F1 & P & R & F1& P & R & F1 & P & R & F1\\
    \midrule
    
    \shortname & \underline{45.1} & 71.8 & 55.4 & 24.8 & \textbf{41.8} & \underline{31.1} & \textbf{52.1} & \underline{66.8} & \textbf{58.5} & \textbf{21.4} & \underline{28.4} & \textbf{24.4} & \underline{55.6} & \textbf{59.5} & \underline{57.5} & \underline{31.4} & \textbf{35.1} & \textbf{33.2} \\
    \midrule
    \multicolumn{19}{c}{\textit{Replacing the Image Captioner with ...}} \\
    BLIPv2 \cite{li2023blip} & 44.4 & 71.7 & 54.8 & 25.2 & 36.2 & 29.7 & 49.2 & 65.7 & 56.3 & 19.3 & 26.8 & 22.4 & 54.1 & 57.9 & 55.9 & 29.5 & 30.3 & 29.9\\
    GIT \cite{wang2022git} & 44.0 & 71.9 & 54.6 & 25.8 & 38.5 & 30.9 & 49.9 & 65.7 & 56.7 & 19.6 & 28.2 & 23.1 & 53.5 & 57.0 & 55.2 & 29.8 & 31.9 & 30.8 \\ 
    VIT-GPT2 \cite{kumar2022imagecaptioning} & 44.8 & 71.1 & 55.0 & \textbf{26.2} & 38.9 & \textbf{31.3} & 49.1 & 65.7 & 56.2 & 19.5 & 28.1 & 23.0 & 54.2 & 57.9 & 56.0 & \textbf{31.9} & 31.9 & \underline{31.9} \\
    OFA \cite{wang2022ofa} & 44.8 & 71.3 & 55.0 & \underline{26.1} & 36.3 & 30.4 & 49.8 & 65.7 & 56.7 & 19.7 & \textbf{28.5} & 23.3 & 54.6 & 57.9 & 56.2 & 30.4 & 30.8 & 30.6 \\
    \midrule
    \multicolumn{19}{c}{\textit{Replacing the Image Generator with ...}} \\
    SDv2 \cite{rombach2022high} & \textbf{45.4} & \underline{72.0} & \textbf{55.7} & 25.0 & 40.4 & 30.9 & 49.6 & 66.0 & 56.6 & 18.7 & 26.2 & 21.8 & 54.7 & 58.9 & 56.7 & 30.1 & 31.8 & 30.9 \\
    SDv1.5 \cite{rombach2022high} & \textbf{45.4} & 71.6 & \underline{55.6} & 24.6 & 40.2 & 30.6 & 50.1 & \textbf{67.3} & \underline{57.4} & \underline{20.1} & \textbf{28.5} & \underline{23.6} & 54.9 & 58.3 & 56.5 & 31.2 & 32.3 & 31.8 \\ 
    Kandinsky \cite{Kandinsky} & 44.8 & \textbf{72.3} & 55.4 & 24.0 & \underline{41.3} & 30.3& \underline{50.4} & 66.2 & 57.2 & 19.9 & 27.2 & 23.0 & \textbf{56.3} & \textbf{59.5} & \textbf{57.9} & 29.8 & \underline{34.2} & \underline{31.9} \\
    \midrule 
    UniCL \cite{liu2022multimedia} & 49.1 & 59.2 & 53.7 & 27.8 & 34.3 & 30.7  &  54.6 & 60.9 & 57.6 & 16.9 & 13.8 & 15.2 & 44.1 & 67.7 & 53.4 & 24.3 & 22.6 & 23.4\\
  \bottomrule
\end{tabular}
\end{table*}

In order to investigate the effects of different components in \shortname, we create ablated systems by removing each of the components. First, we create two variations in the training strategy. In the \textbf{combined training} baseline, we merge the textual event task and the visual event task as one training set and train the model in one stage without freezing any model parameters. In the \textbf{one-round training} baseline, we separate the training of the textual event task and the visual event task. We freeze the visual encoder when training on real textual data and generated visual data, and freeze the textual encoder when training on real textual data and generated visual data. However, we only apply one stage of training and remove the two later stages. 

Next, in the \textbf{w/o augmentation} baseline, we remove all generated multimodal training data and train the network on unimodal data alone. For example, in textual event mention identification, we train the textual encoder and the classifier; the Adapter is removed as well. Finally, the \textbf{w/o Adapter} ablation retains multimodal training data but removes the Adapter network. The cross-attention scores are computed as cosine similarity. For example, in text mention identification, we compute the cosine similarity between each word $h^{\text{text}}_i$ and the visual image encoding $h^{\text{img}}_{\text{CLS}}$. The similarities scores are normalized and used to compute a convex combination of image features, denoted as $g^{\text{text}}_i$. The concatenation of  $h^{\text{text}}_i$ and $g^{\text{text}}_i$ is used for classification. 

The results are shown in Table \ref{tab:abl_result}. The most interesting finding is that the {w/o augmentation}, unimodal baseline outperforms the simplistic combined training strategy by large margins (up to 12.9\% F1 on multimedia event mentions). This clearly demonstrates the difficulties in training on generated multimodal data and the need for a carefully devised training strategy. Second, the one-round training strategy still falls behind the full-fledged \shortname{}, showing the three-stage strategy to be effective. Additionally, the full-fledged \shortname{} appears superior to unimodal training, surpassing by 4.6\% and 4.4\% on multimedia event mention  and argument role extraction respectively. Finally, \shortname{} outperforms the network without Adapter, indicating the advantage of the Adapter design.


\subsection{Choice of Generative Models}

We test if \shortname{} can work with other large pretrained generative models. By default, \shortname{} leverages Stable Diffusion v2.1 \cite{rombach2022high} as the image generator and BLIP \cite{li2022blip} as image captioning model. In this experiment, we test out three different image generators, including Stable Diffusion v1.5 and v2 (SDv1.5 and SDv2) and the Kandinsky model \cite{Kandinsky}. For image captioning, we attempt BLIPv2 \cite{li2023blip}, GIT \cite{wang2022git}, OFA \cite{wang2022ofa}, and VIT-GPT2 \cite{kumar2022imagecaptioning}. 

Table \ref{tab:gen} shows the results. We observe that, while the default settings works well, it often does not achieve the best F1 scores compared to other combinations. In addition, many model combinations outperform UniCL, the previous SOTA model. This demonstrates the generality of the \shortname{} technique.

\section{Conclusions}
\label{Conclusions}

In this paper, we study the problem of multimedia event extraction and investigate the use of image generative networks and image captioning networks to complement existing unimodal training data. The automatically generated multimodal data often contain noise, such as inconsistency with the event label, hallucination, unnatural image artifacts, creating challenges for training. We propose a network, \shortname{}, and a specialized training strategy to cope with augmented multimodal training data. \shortname{} surpasses he prior SOTA by 4.2\% F1 on event mention identification and by 9.8\% F1 on argument identification. An ablation study shows that the design of network structure, the shared adapter, and the iterative training strategy in our method significantly improve performance. We also test the generality of the benefits of our approach to other cross-modality generative models.


\begin{acks}
    This work has been supported by the Nanyang Associate Professorship and the National Research Foundation Fellowship (NRF-NRFF13-2021-0006), Singapore. Any opinions, findings, conclusions, or recommendations expressed in this material are those of the authors and do not reflect the views of the funding agencies.
\end{acks}

\bibliographystyle{ACM-Reference-Format}
\balance
\bibliography{paper_references}










\end{document}